\documentclass[peerreview 10pt,conference]{IEEEtran}
\usepackage[pdftex]{graphicx}
\usepackage[usenames,dvipsnames,table,xcdraw]{xcolor}
\usepackage{array}
\usepackage{multirow}
\usepackage[normalem]{ulem}
\useunder{\uline}{\ul}{}
\usepackage{booktabs}
\usepackage{supertabular,booktabs}
\usepackage{longtable}
\usepackage{lscape}
\usepackage{cite}
\usepackage{url}
\usepackage{hyperref}
\usepackage{dirtytalk}
\usepackage{graphicx}
\usepackage{comment}
\usepackage{makecell}
\usepackage{svg}
\usepackage{amsmath}

\usepackage[caption=false]{subfig}
\graphicspath{{Images/}}
\usepackage[normalem]{ulem}
\useunder{\uline}{\ul}{}

\usepackage{tabularx}
\usepackage{balance}
\usepackage{nameref}

\begin{document}

\title{Hidden Figures in Software Engineering: A Replication Study Exploring Undergraduate Software Students' Awareness of Distinguished Scientists from Underrepresented Groups}

\author{
\IEEEauthorblockN{Ronnie de Souza Santos}
\IEEEauthorblockA{University of Calgary\\
Calgary, AB, Canada \\
ronnie.desouzasantos@ucalgary.ca} 
\and

\IEEEauthorblockN{Italo Santos}
\IEEEauthorblockA{Northern Arizona University\\
   Flagstaff, AZ, US \\
   ids37@nau.edu}
\and

\IEEEauthorblockN{Robson Santos}
\IEEEauthorblockA{UFPE\\
   Recife, PE, Brazil \\
  rtss@cin.ufpe.br}
\and

\IEEEauthorblockN{Cleyton Magalhaes}
\IEEEauthorblockA{UFRPE\\
 Recife, PE, Brazil \\
  cleyton.vanut@ufrpe.br}
}


\IEEEtitleabstractindextext{%
\begin{abstract}
Technology is a cornerstone of modern life, yet the software engineering field struggles to reflect the diversity of contemporary society. This lack of diversity and inclusivity within the software industry can be traced back to limited representation in software engineering academic settings, where students from underrepresented groups are often stigmatized despite the field’s rich history of contributions from scientists from diverse backgrounds. Over the years, studies have revealed that women, LGBTQIA+ individuals, and Black students frequently encounter unwelcoming environments in software engineering programs. However, similar to other fields, increasing awareness of notable individuals from marginalized backgrounds could inspire students and foster a more inclusive environment. This study reports the findings from a replicated global survey with undergraduate software engineering students, exploring their knowledge of distinguished scientists from underrepresented groups. These findings show that students have limited awareness of these figures and their contributions, highlighting the need to improve diversity awareness and develop educational practices that celebrate the achievements of historically marginalized groups in software engineering.
\end{abstract}

}

\maketitle

\IEEEdisplaynontitleabstractindextext

\IEEEpeerreviewmaketitle

\section{Introduction} \label{sec:introduction}
Technology shapes many aspects of contemporary life, influencing work, education, politics, and leisure. However, despite the multicultural nature of modern society, software engineering—the foundation of technological advancement—still lacks diversity in both academic and industrial settings. This limited diversity poses risks that can affect individuals across society~\cite{albusays2021diversity, silveira2019systematic, rodriguez2021perceived}.

Biases in computational algorithms, especially those using machine learning, often disadvantage Black communities by limiting access to services and perpetuating economic disparities, as seen historically in practices like redlining~\cite{lee2018detecting, santos2023perspective}. This issue extends to the legal system, where biases in sentencing and policing disproportionately impact communities of color, deepening social inequalities. Additionally, digital filters on social media reinforce societal beauty standards by lightening skin tones~\cite{marks2019algorithmic, fountain2022moon, owens2020those, lee2018detecting}. In software engineering, gender disparity also highlights the effects of limited diversity, as women frequently encounter hostile, sexist work environments where discrimination and biased evaluations hinder professional growth and reinforce exclusion~\cite{wang2017diversity, sax2017diversifying, zolduoarrati2021value, oliveira2023navigating}.

In software engineering education, similar patterns appear. Students from underrepresented backgrounds often face unwelcoming environments in these programs, which affects their sense of belonging~\cite{ibe2018reflections, ouhbi2020software}. Factors like insufficient representation, lack of support networks, and biases in course content and classroom dynamics contribute to these challenges. Instances of implicit or explicit discrimination based on gender, race, sexual orientation, or other identities can create hostile learning environments~\cite{ibe2018reflections, sax2017diversifying, cheryan2013stereotypical, de2023lgbtqia+}.

The diversity gap in software engineering is particularly puzzling given the field’s history and the influential scientists from underrepresented groups who helped shape it~\cite{haigh2015innovators, gaule2016}. The world’s first programmer was a woman; a gay scientist laid the groundwork for modern computing; and a Black innovator pioneered real-time collaboration, a concept still widely used today. These individuals represent only a few among many from marginalized backgrounds whose contributions have significantly transformed technology.

Given the persistent lack of diversity awareness from academia through to industry and studies showing the exclusion faced by underrepresented students in software engineering~\cite{zolduoarrati2021value, richard2022lgbtq}, a recent study examined how familiar software engineering undergraduate students are with influential figures from underrepresented groups in the field~\cite{de2023diversity}. Results revealed that students receive insufficient classroom exposure to many renowned individuals from these groups—especially Black and LGBTQIA+ pioneers—whose contributions were foundational to today's technologies.

In this study, we replicated the original survey with modifications to increase the breadth and depth of responses. We adopted a different sampling strategy, expanded the questionnaire to include additional scientists from underrepresented groups, and sought to gather data from a broader array of contexts concerning geographical location and student backgrounds. Based on the original study, our goal is to answer the following research question:

\smallskip
\smallskip
\begin{center}
    \fbox{\parbox{0.9\linewidth}{\narrower \noindent \textit{\textbf{RQ:} What do software engineering undergraduate students know about distinguished scientists from underrepresented groups in software engineering and their contributions to the field?} \par}}
\end{center}
\smallskip
\smallskip

Understanding students' familiarity with scientists from underrepresented groups is important for several reasons, including:

\begin{itemize}
    \item \textbf{Inspiration and Motivation}: Awareness of pioneering scientists from diverse backgrounds can inspire students, especially those from similar backgrounds, to continue in software engineering and related fields.
    \item \textbf{Promoting Diversity Awareness}: Understanding students' familiarity with these scientists can guide diversity initiatives in software engineering education, using role models to foster belonging, inclusivity, and representation.
    \item \textbf{Building Evidence for Solutions}: Collecting evidence on diversity challenges in software engineering is important for developing effective approaches to address these issues.
\end{itemize}

This study is organized as follows from this introduction: In Section 2, we provide a background on the history of underrepresented groups in technology. Section 3 summarizes the main aspects of the original study we are replicating in this research. Section 4 outlines the design and implementation of our survey. Section 5 presents the results obtained, followed by a discussion in Section 6. Finally, Section 7 focuses on our conclusions.

\section{Scientists from Underrepresented Groups in the History of Technology} \label{sec:background}

Throughout history, humanity has devised numerous mechanisms and tools for computing. Manual arithmetic, abacuses, and analog machines existed long before the emergence of what we now recognize as computers. Initial strides toward developing the first computer models began in the late 19th century. Since then, numerous pioneering scientists have played crucial roles in transforming computers into indispensable tools supporting our modern society~\cite{dodig2001history}.

While contemporary software engineering often lacks diversity, primarily comprising heterosexual White men~\cite{albusays2021diversity}, the historical landscape of technology illustrates a different narrative. Pioneers and groundbreaking scientists in technology emerge from diverse backgrounds, encompassing various cultures, regions, and educational experiences~\cite{dodig2001history}. Over the decades, individuals from underrepresented groups in the field, including women, Black people, LGBTQIA+ individuals, and many others, have made significant contributions to the evolution of computers and software.

Ada Lovelace and Grace Hopper are notable women in computer science~\cite{freeman1995tap}. Lovelace is celebrated as the world's first programmer~\cite{haigh2015innovators}. In contrast, Hopper, among the early modern computer programmers, made significant contributions to compilers and laid foundational principles for contemporary software testing~\cite{menendez2018grace}. It is noteworthy that historically, programming was initially regarded as a female-dominated occupation, while hardware design was largely male-centric~\cite{albusays2021diversity}.

The LGBTQIA+ community has also left a memorable mark on technological progress. Alan Turing is recognized as the father of computer science and is credited with creating the precursor to modern computers and pioneering fundamental concepts in artificial intelligence~\cite{mijwel2015history}. Christopher Strachey is acknowledged as a programming languages pioneer and played a crucial role in developing the C programming language and early video game development~\cite{campbell1985christopher}. Danielle Bunten Berry, a transgender woman, was a pioneer game developer who created one of the first successful multiplayer games. Additionally, Peter Landin developed concepts that enabled programming languages to transcend specific computer architectures, pioneering the functional programming paradigm~\cite{bornat2009peter}.

Among Black individuals, Clarence Ellis stands as the first African-American to earn a Ph.D. in computer science, holding a significant place in computer science history as he pioneered computer-supported cooperative work, contributing to developing tools and techniques for real-time collaborative document editing~\cite{ellis2022clarence}. Annie J. Easley, one of NASA's initial Black employees, conducted pioneering work on algorithms analyzing alternative power technologies to address energy consumption issues. Additionally, Katherine Johnson and Dorothy Vaughan, two remarkable Black women, made substantial technological contributions at NASA~\cite{shetterly2018hidden}. Frank Greene, the first Black cadet to complete the U.S. Air Force program, contributed to advancing high-speed semiconductor computer-memory systems.

The history of technology is enriched by talented individuals from underrepresented groups who have made significant contributions to technological progress. Today, these individuals persist in driving innovative solutions that shape and benefit our society. Among contemporary scientists, notable figures include Farida Bedwei, a Ghanaian software engineer living with cerebral palsy~\cite{adom2022cultural}, who actively advocates for inclusion within software engineering. Christine M’Lot, an Indigenous educator, creatively integrates computer science education with music~\cite{laskaris2023indigenous}. Furthermore, Lynne Conway, a transgender computer scientist, is renowned for pioneering methods simplifying the design and fabrication of complex microchips~\cite{mead1980introduction}. As software engineering embraces diversity and inclusion, this list of influential figures expands, reflecting the growing contributions from diverse backgrounds.

\section{Original Study} 
\label{sec:original}

The original study~\cite{de2023diversity} gathered data from 128 undergraduate students across four countries, predominantly from Brazil (112 students). Findings revealed that, while students were familiar with certain software engineering figures from underrepresented groups, such as Alan Turing, Ada Lovelace, and Grace Hopper, their knowledge of scientists from other minority groups, particularly Black and Afro-descendant scientists, remained limited. It was anticipated that professors could play a key role in broadening students' understanding by incorporating more information about notable scientists from diverse backgrounds.

Additionally, the study identified primary sources through which students learned about scientists from underrepresented groups. Sources varied by figure: Ada Lovelace was primarily learned about through lectures and social media; Alan Turing through lectures and movies; Christopher Strachey through lectures and social media; Clarence Ellis through social media and research; Grace Hopper through lectures and research; and Peter Landin through lectures and social media.

These outcomes highlight the need to strengthen the teaching of computer science history within undergraduate courses, as students currently engage in limited discussions about numerous distinguished individuals from underrepresented groups who were instrumental in advancing technologies fundamental to software engineering, particularly Black people and LGBTQIA+ individuals. The study emphasized the role of educators in transforming this landscape, suggesting strategies to increase the visibility of underrepresented groups in software engineering, thereby improving diversity awareness and fostering a stronger sense of belonging among students.

\section{Method} 
\label{sec:method}

In this study, the original survey~\cite {de2023diversity} was replicated with strategies to engage participants from multiple countries and slight modifications to the questionnaire, expanding it to include a broader group of scientists from underrepresented groups. The decision to replicate the survey was motivated by the future work recommendations in the original study, which highlighted the need to validate and extend its findings to ensure that the insights gained are applicable across diverse contexts.

Often, a replication study involves the intentional and systematic repetition of previous research, primarily to validate, extend, or reassess previous findings and generalize results across different populations~\cite{shepperd2018role, de2015investigations}. This study represents an exact replication, which repeats a prior study within the same research framework, utilizing identical methods and procedures. Internal replications aim to confirm the reliability of original findings and delve deeper into the outcomes observed~\cite{de2015investigations, bezerra2015replication}.

In this replication, we conducted a cross-sectional survey~\cite {easterbrook2008selecting, pfleeger2001principles, linaker2015guidelines}, adhering to established software engineering guidelines~\cite{ralph2020empirical}, to explore undergraduate software engineering students' familiarity worldwide with key historical figures in technology whose contributions are foundational to their current academic studies. The survey questionnaire and process were refined to improve upon the previous data collection approach from the original study. Our strategy integrated referral sampling, snowball sampling, and the use of Prolific~\footnote{www.prolific.com} to distribute the questionnaire globally among students. Quantitative techniques were then applied to analyze and synthesize the collected data. The detailed methodology is outlined below.

\subsection{Questionnaire} 
\label{sec:questionnaire}

We designed an anonymous questionnaire with the following main sections to gather comprehensive data on students' familiarity with scientists from underrepresented groups in software engineering:

\textbf{Section I: Demographic Information (Part 1).} This section collected essential demographic information from participants, including their country of residence, academic program, and year of enrollment. Since this study aims to capture a broad perspective on students' awareness, participants were recruited not only from software engineering programs but also from related fields, such as computer science, computer engineering, and information systems, as long as these programs included software engineering content in their curriculum. This approach allowed for a diverse sample, reflecting a range of educational backgrounds. 

\textbf{Section III: Open-Ended Familiarity Questions.} This section contained open-ended questions inviting students to describe their knowledge of influential software engineering and technology scientists. This format allowed participants to share their awareness and understanding without predefined response options, providing qualitative insights into students’ familiarity with key figures in the field. These responses help assess general awareness and provide context for the more targeted questions in later sections. 

\textbf{Section IV: Hidden Figures in Software Engineering.} At the core of the questionnaire, this section assessed students' familiarity with specific scientists from underrepresented backgrounds who have made significant contributions to software engineering. Scientists were selected based on three criteria: 1) identification with an underrepresented group in software engineering (such as Black, Hispanic, Indigenous, LGBTQIA+, or individuals with disabilities); 2) notable contributions that are commonly recognized within foundational software engineering courses; and 3) historical status, limited to figures who are no longer living. The list included influential figures such as Ada Lovelace, Alan Turing, Christopher Strachey, Clarence Ellis, Danielle Bunten Berry, Grace Hopper, and Peter Landin.

Ada Lovelace, recognized as the first computer programmer, is often associated with foundational programming courses, where her contributions to early algorithmic thought are highlighted. Alan Turing, known for his work on the theoretical underpinnings of computing, aligns with courses on computer science theory and algorithms. Christopher Strachey, who pioneered early work in programming language concepts, is relevant to programming languages and compiler courses. Clarence Ellis, a pioneer in Computer Supported Cooperative Work, connects to courses on collaborative systems and software development practices. Danielle Bunten Berry, known for her innovative work in multiplayer video games, is pertinent to courses on human-computer interaction and game development. Grace Hopper, who developed one of the first compilers, relates to compiler design and programming languages courses, where her work in testing and debugging is studied. Finally, Peter Landin, a pioneer in the functional programming paradigm, connects to programming paradigms and advanced software development courses, highlighting his impact on language design and functional programming principles.

To allow for comparison, two well-known scientists outside underrepresented groups—Dennis Ritchie and John von Neumann—were also included to enable cross-referencing of our results. Furthermore, we introduced two fictitious detractors, i.e., non-existing scientists, to assess the students' commitment to responding to the survey (Amy Bing and Leonard Cooper). For each scientist, students were asked to indicate how they became familiar with the individual, selecting from sources such as course lectures, assignments, movies, social media, research articles, books, or personal conversations. Students could also note if they could not recall the source of their familiarity or were entirely unfamiliar with the scientist. 

\textbf{Section V: Demographic Information (Part 2).} The final section collected additional demographic details about participants' personal backgrounds, including gender, ethnic group, and sexual orientation. This data supports a nuanced analysis of diversity and inclusivity within the responses, enabling exploration of potential correlations between personal backgrounds and familiarity with underrepresented scientists. To respect participants’ privacy, options were provided to skip any questions they preferred not to answer.

To ensure the validity of the questionnaire, a pilot study was conducted with three software engineering researchers from underrepresented groups, each specializing in equity, diversity, and inclusion (EDI) within software engineering and software engineering education. Their feedback helped refine the questionnaire for clarity, relevance, and accuracy. The questionnaire is available in Table~\ref{tab:survey}. Due to space restrictions, we have not included specific questions regarding the detractors and the scientists who do not belong to underrepresented groups.

\begin{table}[h!]
\caption{Survey Questionnaire}
\label{tab:survey}
\centering
\scriptsize
\begin{tabular}{p{8cm}}
\hline \vspace{0.05cm}
\textbf{ABOUT YOUR UNDERGRADUATE COURSE} \\

1. In which country is your university located? \\

2. What is your course? \\

3. What is your current year of study? \\
\hline
\vspace{0.05cm} 

\textbf{TECHNOLOGY SCIENTISTS THROUGHOUT HISTORY} \\
1. Name up to 3 famous people or influencers that you know in the area of computer/technology. \\

2. Name up to 3 important scientists in the history of computers/technology. \\

3. Underrepresented groups in computer science and technology include women, Black, Hispanics, Indigenous, people with disabilities, and LGBTQ+ individuals. List any scientist that you know who made an important contribution to computer science, software engineering, or technology in general and belongs to an underrepresented group. \\

\hline \vspace{0.05cm} 

\textbf{ABOUT TECHNOLOGY SCIENTISTS THROUGHOUT HISTORY} \\

Select below what you know about each featured scientist. Remember, there are no right or wrong answers. We just want to know what you have heard about them before.\\

1. \textbf{Ada Lovelace} \\
\hspace{0.2cm}( ) Wrote the first computer program in the world. \\
\hspace{0.2cm}( ) Predicted that computers could do more than just crunch numbers. \\
\hspace{0.2cm}( ) Has a programming language named after her. \\
\hspace{0.2cm}( ) Had her contributions recognized only one century after her death. \\
\hspace{0.2cm}( ) I don't know any facts. \\

2. \textbf{Alan Turing} \\
\hspace{0.2cm}( ) Designed the basis of the modern computer. \\
\hspace{0.2cm}( ) Created the algorithm that saved millions of lives during World War II. \\
\hspace{0.2cm}( ) Developed the first substantial study on artificial intelligence. \\
\hspace{0.2cm}( ) Was persecuted for being gay. \\
\hspace{0.2cm}( ) I don't know any facts. \\

3. \textbf{Christopher Strachey} \\
\hspace{0.2cm}( ) Is usually recognized as the first developer of a video game. \\
\hspace{0.2cm}( ) Created fundamental concepts for the C programming language. \\
\hspace{0.2cm}( ) Developed the first research on time-shared computers. \\
\hspace{0.2cm}( ) Identified as a gay man. \\
\hspace{0.2cm}( ) I don't know any facts. \\

4. \textbf{Clarence Ellis} \\
\hspace{0.2cm}( ) Was the first African-American to earn a Ph.D. in Computer Science. \\
\hspace{0.2cm}( ) Was a pioneer in Computer Supported Cooperative Work. \\
\hspace{0.2cm}( ) Developed techniques for real-time collaborative editing of documents. \\
\hspace{0.2cm}( ) Faced a non-inclusive environment as an African-American student. \\
\hspace{0.2cm}( ) I don't know any facts. \\

5. \textbf{Danielle Bunten Berry} \\
\hspace{0.2cm}( ) Was a pioneer in developing multiplayer video games. \\
\hspace{0.2cm}( ) Her 1983 game M.U.L.E. was listed among the best games of all time. \\
\hspace{0.2cm}( ) Identified as a transgender woman. \\
\hspace{0.2cm}( ) Received a lifetime achievement award from the Game Developers Association. \\
\hspace{0.2cm}( ) I don't know any facts. \\

6. \textbf{Grace Hopper} \\
\hspace{0.2cm}( ) Reported the first computer bug. \\
\hspace{0.2cm}( ) Designed one of the first compilers. \\
\hspace{0.2cm}( ) Developed the COBOL programming language. \\
\hspace{0.2cm}( ) Was the first woman to earn a Ph.D. in Mathematics at Yale University. \\
\hspace{0.2cm}( ) I don't know any facts. \\

7. \textbf{Peter Landin} \\
\hspace{0.2cm}( ) Created concepts enabling programming languages to run on any computer. \\
\hspace{0.2cm}( ) Was a pioneer in the functional programming paradigm. \\
\hspace{0.2cm}( ) Advocated for LGBTQIA+ rights. \\
\hspace{0.2cm}( ) Identified as a bisexual man. \\
\hspace{0.2cm}( ) I don't know any facts. \\

\\ 
8. Tell us how you heard about each scientist, including if it was through a course or lecture, an assignment, a movie or video, research or a book, a blog or post on social media, colleagues and peers at school, family and friends, or if you do not remember or haven’t heard anything until now. \\
\hline
\vspace{0.05cm} 

\textbf{ABOUT YOU} \\
1. What ethnic group do you identify with? \\
2. What best describes your gender? \\
3. Do you identify as transgender? \\
4. What is your sexual orientation? \\
\hline
\end{tabular}
\end{table}

\subsection{Participants} \label{sec:participants}

Determining the precise population of software engineering students globally presents a significant challenge due to several factors. Firstly, the definition and scope of software engineering programs vary across educational institutions and countries. Some institutions might offer dedicated software engineering courses, while others may incorporate software engineering modules into broader computer science or engineering programs. Additionally, variations in terminology and program structures make it difficult to define the population precisely. Hence, achieving a fully representative sample across different regions and educational systems globally posed challenges stemming from limited access to certain student cohorts, language barriers, and differential levels of participation across countries and institutions. In this sense, our population encompassed undergraduate students from various countries enrolled in university programs that focus on software engineering at some level and who were willing to participate in the study.

\subsection{Data Collection} \label{sec:collection}

We utilized different non-probability sampling methods to collect data for our study. Initially, following the original study we are replicating, we employed convenience sampling by contacting software engineering professors from different universities. These professionals were identified from software engineering conference program committees. We requested their cooperation in distributing our questionnaire to undergraduate students within their universities using email lists and other communication channels available. Following this, we implemented a snowball sampling technique, encouraging participants to share our questionnaire with other interested students and expanding the study's reach through participant referrals. 

Although these methods successfully gathered data from a relatively large participant pool previously, their limitations resulted in a lack of geographical diversity within our sample. For instance, a substantial number of participants originated from a single country. To address this limitation, we employed the Prolific platform to support participant recruitment alongside convenience and snowball sampling in this study. Prolific is a widely-utilized crowdsourcing platform with over 150,000 active users~\cite{russo2022recruiting}, extensively employed for conducting software engineering surveys across diverse contexts, encompassing professionals and students. This strategic adjustment broadened our participant pool and ensured a more comprehensive representation across various geographical regions.

Combining different data collection strategies increased the likelihood of including participants outside our target population. To manage this, our data collection process included several filters and screening questions to verify the authenticity of participants' responses and prevent inaccurate or random answers. We introduced a pre-screening validation section with recommended questions to assess students' programming skills \cite{danilova2021you}. Given that software engineering students are expected to possess fundamental knowledge in programming, such as variable types, this section aimed to ensure participants belonged to the field. In the Prolific platform, access to the complete survey questionnaire was granted only after completing this pre-screening validation section. Furthermore, detractors were integrated into our questionnaire to enhance the quality of the filter. By introducing fictitious scientists and identifying participants who claimed familiarity with these invented individuals, we could exclude those respondents who provided random answers from our final sample. Data collection was concluded with 89 participants coming from 20 different countries.

\subsection{Data Analysis} \label{sec:analysis}

In this study, both quantitative and qualitative data were collected, with most qualitative data being quantifiable. Therefore, thematic analysis~\cite{Cruz2011} was employed to convert qualitative data from open-ended survey questions into categories suitable for quantitative analysis. We organized the data into structured groups after reviewing and coding responses based on recurring themes. 

Descriptive statistics~\cite{george2018descriptive} were then applied to examine our sample's main characteristics and address the research questions. We summarize essential features, characteristics, and patterns within our dataset, encompassing measures of central tendency (mean, median, mode), variability (range, variance, standard deviation), and distribution representations (e.g., histograms and frequency distributions)~\cite{george2018descriptive}. In survey analysis, these methods provide a structured overview of collected data.

Our analysis involved segmenting participants' responses into subgroups using statistical measures like means, proportions, totals, and ratios to gain a comprehensive understanding of the data. Through these aggregations, we also explored correlations between students' backgrounds, experiences, and knowledge of computer scientists from underrepresented groups.

\subsection{Ethics} 

Following ethical guidelines, no personal information regarding the participants (such as names, emails, or universities) was gathered during this study to uphold participant anonymity. Before starting the questionnaire, participants were briefed about the study's objectives and were required to consent to use their responses for scientific purposes.

\section{Findings} 
\label{sec:findings}

As detailed in Section~\ref{sec:collection}, we utilized a combination of convenience sampling, snowball sampling, and crowdsourcing to gather data from a broad spectrum of software engineering students worldwide. This method yielded a participant pool of 89 individuals representing 20 countries. Our participant sample embodies diversity across various dimensions, encompassing differences in the software engineering programs they are enrolled in, their academic year, and the diverse motivations influencing their choice to pursue a career in software engineering. Furthermore, our sample demonstrates diversity in terms of gender, ethnicity, and sexual orientation. This diversity is fundamental to our study, which focuses on understanding how students from various groups recognize scientists from underrepresented backgrounds in the history of software engineering. A summary of our sample demographics can be found in Table~\ref{tab:Demographics}.

\begin{table}
\centering
\scriptsize
\caption{Demographics}
\renewcommand{\arraystretch}{1}
\label{tab:Demographics}
\begin{tabular}{llr}
\hline\noalign{\smallskip}
\multirow{3}{*}{ \textbf{Gender} } 
& Men & 64\\
& Women & 23$^a$ \\
& Non-Binary & 2 \\ \midrule 

\multirow{5}{*}{ \textbf{Ethnicity} } 
& White & 47 \\
& Black & 15 \\
& Hispanic & 13 \\
& Asian & 12 \\
& Mixed Ethnicity & 2 \\ \midrule 

\multirow{8}{*}{ \textbf{Sexual Orientation} } 
& Straight & 66 \\
& Bisexual & 10 \\
& Gay & 6 \\
& Lesbian & 2 \\
& Pansexual & 1 \\
& Asexual & 1 \\
& Other & 1 \\
& Prefer not to answer & 2 \\ \midrule 

\multirow{5}{*}{ \textbf{Program Focus} } 
& Computer Science & 55 \\
& Software Engineering & 10 \\
& Information Technology & 8 \\
& Computer Engineering & 7 \\
& Other Software-Related & 9 \\ \midrule 

\multirow{5}{*}{ \textbf{Year} } 
& 1st & 7 \\
& 2nd & 18 \\
& 3rd & 31 \\
& 4th & 19 \\
& 5th & 14 \\ \midrule 

\multirow{20}{*}{ \textbf{Country} }
& Australia & 3 \\
& Austria & 1 \\
& Brazil & 7 \\
& Canada & 7 \\
& Chile & 2 \\
& Estonia & 1 \\
& Germany & 1 \\
& Greece & 2 \\
& Hungary & 3 \\
& Israel & 1 \\
& Italy & 3 \\
& Latvia & 1 \\
& Mexico & 9 \\
& Netherlands & 4 \\
& New Zealand & 1 \\
& Poland & 10 \\
& Portugal & 13 \\
& South Africa & 11 \\
& United Kingdom & 2 \\
& United States & 7 \\

\noalign{\smallskip}\hline
\end{tabular}

\flushleft
\footnotesize{Notes: $^a$19 cisgender, 4 transgender}
\end{table}


\subsection{Distinguished Scientists and Well-Known Public Figures}

In the next stage of our study, we requested students to cite up to three individuals whom they believed had a notable influence or made significant contributions to the technology field. Our objective at this stage was to explore whether students would spontaneously mention scientists, regardless of whether these figures directly influenced their choice to pursue software engineering. As a result, we observed that only 20\% of the sampled students, which accounts for 18 individuals, spontaneously mentioned scientists while listing up to three individuals whom they perceived to substantially influence the technology field. This approach enabled us to gauge the extent of their recognition and distinguish between scientists and various personalities within software engineering. 

Subsequently, we proceeded with a more targeted investigation by prompting students to specifically identify up to three influential scientists from the history of technology whom they recognized. This step aimed to explore further the potential disparity between the historical importance of scientists in technology and their perceived significance within the field of software engineering by students. As a result, 28\% of the sample (25 students) acknowledged that they were not familiar with any scientist who significantly contributed to shaping software engineering. On the other hand, among the participants familiar with at least one scientist in the field, we identified 35 different names spanning various areas within software engineering. Notably, Alan Turing was the most frequently mentioned scientist, receiving 47 mentions, followed by Ada Lovelace with 22 mentions. The list also included John von Neumann (16 mentions), Grace Hopper (11 mentions), Linus Torvalds (6 mentions), Tim Berners-Lee (5 mentions), Edsger Dijkstra (5 mentions), and Charles Babbage (4 mentions).

These findings align with those obtained from the previous study. Once again, Alan Turing and Ada Lovelace emerge as the most mentioned figures, indicating a general awareness among students regarding the contributions of underrepresented groups to technology. However, in general, these findings underscore that although most students are familiar with scientists in the technology field, they may not view these scientists as influential to the subjects they learn in their courses or instrumental in guiding their choice to pursue software engineering. This suggests a potential discrepancy between acknowledging historical and notable figures in the field and recognizing their direct influence on their career.

\subsection{Familiarity with Distinguished Scientists from Underrepresented Groups in Software Engineering}

When students were asked to spontaneously mention scientists from underrepresented groups in technology, 43\% of our sample indicated unfamiliarity with any scientists falling within this category. Despite the question's clarification that this group includes Women, Black people, Hispanics, Indigenous, people with disabilities, and LGBTQ+ individuals, a significant number of participants were unable to name scientists from these backgrounds.

Among the responses, Alan Turing and Ada Lovelace were the most frequently cited scientists, with 18 and 16 mentions, respectively. However, these numbers were significantly lower compared to the count when students were asked about scientists in general. This disparity indicates that many students might lack awareness of these scientists' personal backgrounds, which might limit their ability to recognize similarities between their own backgrounds and those of these scientists.

Additionally, a few other scientists from underrepresented groups were mentioned spontaneously by participants, including Grace Hopper, Katherine Johnson, Lynn Conway, Vint Cerf, Margaret Hamilton, Annie Easley, Dorothy Vaughan, Anita Borg, Mark Dean, Hanna Oktaba, Gonzales Camarena, and Carol Shaw. Some of these scientists met the criteria outlined for the list of scientists presented to the students in the subsequent phase of the survey. Overall, 57\% of our sample, encompassing 51 participants, were able to spontaneously cite at least one scientist from an underrepresented group in software engineering.

When participants were prompted to recognize scientists from the provided list of individuals who met our specific criteria (i.e., being from underrepresented groups within software engineering, having notable contributions taught in fundamental courses across programs, and no longer alive), an overwhelming majority of the sample, over 95\%, indicated having some level of familiarity with these scientists. This implies a general awareness among participants regarding these (\textit{somewhat hidden}) figures within software engineering.

Specifically, 88\% of participants reported familiarity with Alan Turing, followed by 59\% with Ada Lovelace and 46\% with Grace Hopper. However, a smaller percentage recognized Clarence Ellis (4\%) and Christopher Strachey (4\%). Additionally, only 3\% of participants were familiar with Peter Landin. Only one participant demonstrated familiarity with the work of Danielle Bunten Berry at first. These percentages indicate that while participants may have learned about these renowned scientists from underrepresented groups, they might not retain this information, which results in diminishing their perception of these scientists' contributions and, subsequently, the potential impact of diversity in the field of software engineering.

\subsection{What software engineering undergraduate students know about distinguished scientists from underrepresented groups in software engineering and their contributions to the field?}

We presented students with facts about the scientists from underrepresented groups in our list to evaluate their familiarity with these figures beyond their significant contributions to the field. The outcomes (summarized in Table \ref{tab:know}) revealed that while students exhibited extensive knowledge about the professional backgrounds of these scientists, they displayed less familiarity with their personal backgrounds and their status as members of underrepresented groups within software engineering.

Regarding those that are more well-known among students, the same percentage of participants who recognized Turing's foundational role in designing the modern computer were also aware of the persecution he faced due to his sexual orientation. However, Ada Lovelace and Grace Hopper had different levels of awareness. While participants were familiar with some of their contributions to technology, they lacked awareness of the challenges these women encountered within the male-dominated field and their individual accomplishments beyond their work.

As for the lesser-known scientists, our findings highlighted variations in students' familiarity with their contributions versus their personal backgrounds. For instance, students were more familiar with Clarence Ellis's individual accomplishments than his specific contributions to the technologies that are currently studied in software engineering programs. On the other hand, students recognized some of the contributions of Christopher Strachey and Peter Landin but were unaware of their status as members of underrepresented groups within software engineering. Finally, even though only one participant recognized Danielle Bunten Berry in the earlier stages of the survey when presented with facts about her work, a few students recognized some of her contributions, possibly linking the contribution to the name after being introduced to them. Yet, similar to Christopher Strachey and Peter Landin, they were not cognizant of her status as a member of a minority group in the field, specifically as a transgender woman.

\begin{table}
\centering
\scriptsize
\caption{Familiarity and Facts}
\renewcommand{\arraystretch}{0.8}
\label{tab:know}
\begin{tabular}{llr}
\hline
\textbf{Scientist} & \textbf{Participants} \\
\hline

\textbf{Ada Lovelace} \\
\hspace{0.2cm} First computer program & 45 individuals \\
\hspace{0.2cm} Predicted computer applications & 34 individuals \\
\hspace{0.2cm} Programming language named after her & 11 individuals \\
\hspace{0.2cm} Recognized posthumously & 22 individuals \\
\hspace{0.2cm} I don't know any facts & 21 individuals \\
\midrule 

\textbf{Alan Turing} \\
\hspace{0.2cm} Basis of modern computer & 48 individuals \\
\hspace{0.2cm} WWII algorithm saved lives & 53 individuals \\
\hspace{0.2cm} Early study on AI & 28 individuals \\
\hspace{0.2cm} Persecuted for being gay & 49 individuals \\
\hspace{0.2cm} I don't know any facts & 11 individuals \\
\midrule 

\textbf{Christopher Strachey} \\
\hspace{0.2cm} First video game & 10 individuals \\
\hspace{0.2cm} Concepts for C language & 2 individuals \\
\hspace{0.2cm} Time-shared computers research & 7 individuals \\
\hspace{0.2cm} Identified as gay & 2 individuals \\
\hspace{0.2cm} I don't know any facts & 76 individuals \\
\midrule 

\textbf{Clarence Ellis} \\
\hspace{0.2cm} First African-American Ph.D. in CS & 6 individuals \\
\hspace{0.2cm} Pioneer in cooperative work & 4 individuals \\
\hspace{0.2cm} Real-time editing techniques & 2 individuals \\
\hspace{0.2cm} Experienced exclusion in school & 3 individuals \\
\hspace{0.2cm} I don't know any facts & 82 individuals \\
\midrule 

\textbf{Danielle Bunten Berry} \\
\hspace{0.2cm} Multiplayer game pioneer & 3 individuals \\
\hspace{0.2cm} M.U.L.E. game top-rated & 6 individuals \\
\hspace{0.2cm} Identified as transgender & 0 individuals \\
\hspace{0.2cm} Lifetime award from Game Assoc. & 0 individuals \\
\hspace{0.2cm} I don't know any facts & 82 individuals \\
\midrule 

\textbf{Grace Hopper} \\
\hspace{0.2cm} First computer bug reported & 16 individuals \\
\hspace{0.2cm} Early compiler designer & 22 individuals \\
\hspace{0.2cm} Developed COBOL & 22 individuals \\
\hspace{0.2cm} First Math Ph.D. woman at Yale & 8 individuals \\
\hspace{0.2cm} I don't know any facts & 82 individuals \\
\midrule 

\textbf{Peter Landin} \\
\hspace{0.2cm} Cross-platform language concepts & 5 individuals \\
\hspace{0.2cm} Functional programming pioneer & 3 individuals \\
\hspace{0.2cm} Advocated LGBTQIA+ rights & 1 individual \\
\hspace{0.2cm} Identified as bisexual & 1 individual \\
\hspace{0.2cm} I don't know any facts & 83 individuals \\

\hline
\end{tabular}
\end{table}

This discrepancy observed between the capacity of software engineering students to recognize achievements and understand the historical context and challenges faced by these scientists from underrepresented groups is significant in assessing the students' depth of knowledge and perception regarding the role that diversity plays in modern software engineering. In this sense, apart from investigating students' knowledge about scientists from underrepresented groups in software engineering, it is also important to explore the origins of their knowledge about these figures to start formulating effective strategies to enhance awareness about diversity in the field. Therefore, during the final phase of this survey, we specifically inquired students about their sources of information regarding these scientists. The outcomes of this investigation are outlined in Table \ref{tab:source}.

Based on our findings (Table~\ref{tab:source}), lectures, assignments, and research have emerged as consistent methods for disseminating information about scientists from underrepresented groups in software engineering. However, our research highlighted a significant gap in classroom teachings about numerous renowned individuals from underrepresented groups who significantly impacted software engineering. Additionally, while Alan Turing and Ada Lovelace possess more widespread recognition, there exists an abundance of resources available about their lives and achievements, including movies and extensive online materials, such as social media posts. Conversely, information about other notable scientists in the field from underrepresented groups remains relatively limited. Finally, our observations also suggest the potential for informal knowledge sharing among colleagues, peers, and social connections as a promising avenue to enhance students' awareness and understanding of diversity within the field of software engineering.

In summary, answering our research question, the results of this replicated study confirm that software engineering undergraduate students possess varying levels of awareness and knowledge about distinguished scientists from underrepresented groups in their field. There is more familiarity among students with notable figures such as Alan Turing and Ada Lovelace, which can be attributed to the extensive resources available about them. However, there exists a knowledge gap regarding other influential scientists from marginalized backgrounds, particularly individuals from the Black and LGBTQIA+ communities, whose significant contributions to software engineering remain relatively less explored in classroom lessons. Overall, there is room for improvement in their knowledge and recognition of these influential individuals and their contributions to the field, aiming to increase diversity awareness among students.

\begin{table}
\centering
\scriptsize
\caption{Source of Information}
\renewcommand{\arraystretch}{1}
\label{tab:source}
\begin{tabular}{llr}
\hline
\textbf{Scientist} & \textbf{Source} & \textbf{Participants} \\
\hline
\multirow{3}{*}{ \textbf{Ada Lovelace} } 
& Assignment & 5 individuals \\
& Blog or Social Media & 7 individuals \\
& Colleagues and Peers & 7 individuals \\
& Course or Lecture & 35 individuals \\
& Family and Friends & 2 individuals \\
& Movie or Video & 7 individuals \\
& Research & 15 individuals \\
& I do not remember where & 11 individuals \\
\midrule 

\multirow{3}{*}{ \textbf{Alan Turing} } 
& Assignment & 14 individuals \\
& Blog or Social Media & 7 individuals \\
& Colleagues and Peers & 12 individuals \\
& Course or Lecture & 53 individuals \\
& Family and Friends & 6 individuals \\
& Movie or Video & 36 individuals \\
& Research & 22 individuals \\
& I do not remember where & 4 individuals \\
\midrule 

\multirow{3}{*}{ \textbf{Christopher Strachey} } 
& Assignment & 2 individuals \\
& Blog or Social Media & 1 individual \\
& Colleagues and Peers & 2 individuals \\
& Research & 4 individuals \\
& I do not remember where & 5 individuals \\
\midrule 

\multirow{3}{*}{ \textbf{Clarence Ellis} } 
& Assignment & 1 individuals \\
& Blog or Social Media & 2 individuals \\
& Colleagues and Peers & 1 individuals \\
& Course or Lecture & 3 individuals \\
& Movie or Video & 1 individuals \\
& Research & 3 individuals \\
\midrule 

\multirow{3}{*}{ \textbf{Danielle Bunten Berry} } 
& Colleagues and Peers & 1 individuals \\
& Course or Lecture & 1 individuals \\
& Family and Friends & 1 individuals \\
& Movie or Video & 2 individuals \\
& Research & 2 individuals \\
\midrule 

\multirow{3}{*}{ \textbf{Grace Hopper} } 
& Assignment & 7 individuals \\
& Blog or Social Media & 7 individuals \\
& Colleagues and Peers & 5 individuals \\
& Course or Lecture & 12 individuals \\
& Movie or Video & 7 individuals \\
& Research & 16 individuals \\
& I do not remember where & 11 individuals \\
\midrule 

\multirow{3}{*}{ \textbf{Peter Landin} } 
& Course or Lecture & 1 individuals \\
& Family and Friends & 1 individuals \\
& Movie or Video & 1 individuals \\
& Research & 2 individuals \\
& I do not remember where & 2 individuals \\


\noalign{\smallskip}\hline
\end{tabular}
\flushleft
\end{table}


\subsection{Correlations Between Student's Backgrounds and their Familiarity with Scientists}

The recognition of scientists within the dataset reveals interesting patterns, particularly with respect to gender, ethnicity, and sexual orientation. Alan Turing stands out as a universally recognized figure, transcending demographic categories. His foundational work in algorithms has made him widely known and highly recognized across the whole sample. Though Turing’s legacy as an LGBTQ+ figure in computer science might suggest higher familiarity among LGBTQ+ students, his recognition across all groups reflects his broad impact on the field. 

Among the prominent female scientists in technology, Ada Lovelace and Grace Hopper are widely recognized, especially by female and non-binary participants. Ada Lovelace has a high recognition rate among female students (74\%) and is universally recognized among non-binary participants (100\%). Similarly, Grace Hopper is recognized by 61\% of female students and all non-binary students, compared to a lower rate of 39\% among males. Additionally, both Lovelace and Hopper show slightly higher recognition among LGBTQ+ orientations. These patterns suggest a connection between gender identity and the recognition of female figures in software engineering, where female, non-binary, and LGBTQ+ students may find greater representation and inspiration in these figures.

Clarence Ellis, a pioneering Black computer scientist, exhibits a unique recognition pattern. He is primarily recognized by Black/African American and Hispanic/Latinx participants, indicating resonance within minority groups likely due to his status as one of the first Black computer scientists. This finding reflects the broader theme of representation, where Ellis's contributions may hold particular significance for students from racial minority backgrounds.

Danielle Bunten Berry, Peter Landin, and Christopher Strachey, all notable LGBTQ+ figures, have very low recognition across all demographics. Danielle Bunten Berry was recognized only by one Black/African American male student, while Peter Landin has familiarity among bisexual Hispanic/Latinx participants. Christopher Strachey is primarily recognized by Black/African American and Hispanic/Latinx males, though his visibility remains low compared to other figures. The modest recognition, despite their contributions to foundational areas in programming, suggests a potential gap in awareness around LGBTQ+ pioneers, highlighting the need for greater visibility of diverse contributors within software engineering.

Finally, it is important to highlight that widely recognized scientists who are not part of underrepresented groups, such as John von Neumann and Dennis Ritchie, show demographic alignment with traditional technical groups. John von Neumann is particularly well-known among male and heterosexual participants, as is Dennis Ritchie. Ritchie’s recognition is higher among White/Caucasian and Hispanic/Latinx males, with lower familiarity among females and LGBTQ+ groups. These patterns emphasize a gender gap in recognizing figures associated with programming languages and computer architecture.

\section{Discussions}  \label{sec:discussions}

In general, our findings indicate that students from more diverse backgrounds tended to recognize a broader range of scientists from underrepresented groups. For example, students identifying with minority groups showed increased awareness of the personal identities of scientists, such as the LGBTQIA+ background of Christopher Strachey or the pioneering work of Clarence Ellis as an African-American. Conversely, students from less diverse backgrounds were primarily familiar with widely recognized figures like Alan Turing or Ada Lovelace, often overlooking scientists who made critical yet less-publicized contributions. These findings suggest that students’ personal experiences and EDI exposure play a significant role in shaping their awareness of diversity within software engineering and highlight the need for inclusive curricula to bridge these familiarity gaps.

\subsection{Comparing Studies: Original and Replication}

We compared the findings from the previous study~\cite{de2023diversity} and this replication and identified that both studies share the conclusion that lectures are a primary method of information dissemination about scientists from underrepresented groups, but they also highlight substantial gaps in classroom teachings on the breadth of influential figures within software engineering. Our study confirms that while figures like Alan Turing and Ada Lovelace are widely recognized and benefit from a wealth of available resources—including movies, online articles, and social media content—lesser-known scientists from diverse backgrounds do not enjoy the same level of exposure. The previous study~\cite{de2023diversity} further supports this by identifying that Turing and Lovelace received the highest mentions, while other notable figures like Grace Hopper, Clarence Ellis, and Christopher Strachey remained relatively unknown among students, indicating a shared need to expand the representation of underrepresented groups in educational materials.

In our study, Grace Hopper is particularly recognized by female and non-binary students, suggesting her status as a relatable figure for individuals from diverse gender identities. This complements the previous study’s findings~\cite{de2023diversity}, where Hopper was also recognized but with lower mentions than Turing and Lovelace. These complementary insights emphasize the opportunity for educators to emphasize female role models in computing, as Hopper’s visibility could serve as an important source of inspiration and representation for female and non-binary students in STEM fields.

Finally, our findings highlight that informal avenues, such as peer discussions and social connections, may be valuable in broadening students' awareness of lesser-known scientists from underrepresented groups. The previous study supports this perspective by showing that students often learn about underrepresented scientists through social media and independent research, suggesting that integrating diverse historical figures into informal learning and social channels could address existing gaps. Together, both studies emphasize the importance of expanding both formal curricula and informal knowledge-sharing practices to deepen students' understanding of diversity and the impact of underrepresented groups within software engineering.

\subsection{Enfolding the Literature}

Similar to our findings, the literature from other fields indicates that the underrepresentation of marginalized pioneers in educational materials leads to lower recognition and familiarity among students. In psychology, for example, marginalized pioneers are often absent from textbooks, which affects students’ recognition of these figures and their contributions~\cite{kelly2024making}. This lack of representation could be addressed by incorporating more inclusive curricula. Our study echoes these findings by demonstrating that expanding diversity awareness in software engineering can positively influence students’ understanding of the field’s history and the diversity of its contributors.

Additionally, studies show that students are more likely to connect with and be inspired by role models with whom they share certain characteristics, such as gender identity, as this helps break down initial misconceptions about who can succeed in these fields~\cite{buck2008examining}. This aligns with our observation that female and non-binary students are especially likely to recognize female pioneers like Ada Lovelace and Grace Hopper. In both cases, the presence of relatable figures appears to play a significant role in supporting diversity by offering students from underrepresented groups a sense of belonging and inspiration within their fields.

Finally, previous research has shown that learning about the obstacles faced by pioneering figures helps students view scientists as hardworking individuals who overcame significant barriers, which increases students’ interest in the career~\cite{hong2012learning}. Our findings suggest that this approach is yet to be validated in software engineering, where lesser-known figures from underrepresented groups, such as LGBTQ+ pioneers, often remain unfamiliar to students. We hypothesize that by highlighting the diverse journeys of these pioneers, educators can provide a more comprehensive picture of the field, inspiring students to relate to these figures and supporting a more inclusive academic environment~\cite{hong2012learning, ferguson2021role}.

\subsection{Implications}

The outcomes derived from our survey carry significant implications for educational practices. The evident lack of recognition of scientists from underrepresented groups in software engineering among students demonstrates an opportunity for programs and lecture adaptations. By incorporating information about these overlooked figures into software engineering coursework, educators can instill a more inclusive and diverse learning environment, improving diversity awareness and helping students from marginalized groups to feel represented in the field. A teaching strategy that embraces diversity can actively contribute to breaking stereotypes, promoting diversity, and nurturing an environment that values the contributions of all individuals in the field of technology.

The insights from our study are of significant importance to software engineering researchers. This study pioneers the exploration of role modeling within software engineering education. Researchers can leverage these findings to investigate and refine role-modeling techniques in educational settings. Our study emphasizes the importance of integrating the historical contributions of influential software engineering scientists from underrepresented groups into educational frameworks, and this perspective provides researchers with a basis to develop innovative methodologies and educational strategies aimed at fostering diversity and promoting inclusivity.

Our findings implicate several opportunities for future research in software engineering education. Currently, there is a critical need to develop comprehensive guidelines that can effectively guide educators in fostering a more inclusive environment within software engineering programs. Implementing role modeling methods tailored for software engineering courses could significantly contribute to this goal. Additionally, further research should focus on formulating strategies to tackle equity, diversity, and inclusion issues within highly technical courses, such as those centered on coding or hardware. Finally,  longitudinal studies could help understand the long-term impacts of inclusive teaching practices on students' careers in the software engineering field.

\subsection{Limitations}

Our study acknowledges limitations related to selection and sampling bias, primarily due to the difficulty of accurately identifying the global population of software engineering students. To address this, we utilized convenience sampling, snowball sampling, and crowdsourcing to expand our participant pool. Crowdsourcing, in particular, helped balance the geographical diversity of our sample, addressing the issue seen in prior research where responses were disproportionately from a single country. This diverse sampling aligns with our study’s focus on representing various backgrounds in terms of geography, gender, ethnicity, and sexual orientation.

We also considered potential response biases, given the survey’s focus on underrepresented groups in software engineering. To mitigate this, we designed the survey with detractor questions and included both marginalized and non-marginalized scientists. This approach aimed to filter out participants who might not be fully engaged in the study, thereby enhancing data quality. Our sampling methods helped ensure a wide array of social backgrounds among participants, reducing the likelihood of skewed responses driven by social desirability.

Furthermore, we incorporated open-ended questions before structured alternatives, encouraging participants to express their genuine perspectives without the influence of preset options. This gradual approach allowed participants to explore their motivations for entering software engineering, discuss their inspirations, and eventually reflect on their knowledge of scientists and underrepresented groups in the field. Finally, while the study’s findings may not be strictly generalizable, they offer adaptable insights that can stimulate knowledge exchange, providing valuable context for educators and researchers interested in diversity within software engineering.

\section{Conclusions}  \label{sec:conclusions}

This study investigated the perceptions and awareness of software engineering students regarding distinguished scientists from underrepresented groups in the field. Our primary goal was to understand these students' familiarity with historical contributions made by these scientists to foundational concepts and technologies in software engineering. Additionally, we aimed to explore whether there were connections between students' knowledge of these scientists and their awareness of diversity within the field. 

Through this work, we sought to shed light on the ongoing diversity challenges in software engineering education. To achieve this, we utilized a range of data collection methods, including convenience sampling, snowball sampling, and crowdsourcing techniques. This approach enabled the recruitment of 89 software engineering students from 20 different countries, encompassing a broad spectrum of geographical locations, cultural backgrounds, genders, ethnicities, and sexual orientations. The diversity in our sample provided a valuable foundation for analyzing patterns of awareness and recognition across demographic.

Our findings revealed that while a significant number of students recognized widely known figures like Alan Turing and Ada Lovelace, there was a notable lack of awareness regarding other scientists from underrepresented groups who are less frequently mentioned in lectures or mainstream media. This knowledge gap underscores the need to enhance diversity awareness within software engineering education. In conclusion, our findings demonstrate the need to improve diversity awareness among software engineering students and advocate for inclusive educational strategies that amplify the presence and contributions of historically marginalized groups.

\section{Data Availability}
The data collected through the survey can be accessed via this link: \url{https://figshare.com/s/fb9d4d9e28a6f667ea14?file=50158524}

\ifCLASSOPTIONcaptionsoff
  \newpage
\fi

\balance
\bibliographystyle{IEEEtran}
\bibliography{bib.bib}

\end{document}